\documentclass[letterpaper]{article} 
\usepackage[]{aaai25}  
\usepackage{times}  
\usepackage{helvet}  
\usepackage{courier}  
\usepackage[hyphens]{url}  
\usepackage{graphicx} 
\usepackage{natbib}  
\usepackage{caption} 
\usepackage{array} 
\usepackage{algorithm}
\usepackage{algorithmic}
\usepackage{booktabs}
\usepackage{amsmath}
\usepackage{tcolorbox}
\usepackage{amssymb}  
\usepackage{enumitem}

\usepackage{tikz}
\usetikzlibrary{positioning,arrows,shapes,fit,backgrounds}
\usepackage{forest}
\usetikzlibrary{arrows.meta}
\usepackage{newfloat}
\usepackage{listings}
\DeclareCaptionStyle{ruled}{labelfont=normalfont,labelsep=colon,strut=off} 
\floatstyle{ruled}
\newfloat{listing}{tb}{lst}{}
\floatname{listing}{Listing}
\definecolor{barcolor}{RGB}{74, 111, 165} %
\newcommand{\barwidth}{3cm} %

\usepackage{xcolor}

\usepackage{amssymb} 

\usepackage{soul}
\definecolor{lightorange}{RGB}{255, 228, 196}  
\definecolor{lightblue}{RGB}{217, 232, 255}    
\definecolor{lightyellow}{RGB}{255, 245, 213}  
\definecolor{lightpurple}{RGB}{237, 214, 229}  
\definecolor{lightgreen}{RGB}{223, 244, 222}   
\definecolor{lightred}{RGB}{255, 217, 217}     
\newcommand{\hlorange}[1]{\sethlcolor{lightorange}\hl{#1}}
\newcommand{\hlblue}[1]{\sethlcolor{lightblue}\hl{#1}}
\newcommand{\hlyellow}[1]{\sethlcolor{lightyellow}\hl{#1}}
\newcommand{\hlpurple}[1]{\sethlcolor{lightpurple}\hl{#1}}
\newcommand{\hlgreen}[1]{\sethlcolor{lightgreen}\hl{#1}}
\newcommand{\hlred}[1]{\sethlcolor{lightred}\hl{#1}}

\usepackage{xcolor}         

\title{Whose 
\textcolor[RGB]{0,41,93}{P}%
\textcolor[RGB]{0,60,135}{e}%
\textcolor[RGB]{0,79,178}{r}%
\textcolor[RGB]{0,98,221}{s}%
\textcolor[RGB]{41,117,255}{o}%
\textcolor[RGB]{82,137,255}{n}%
\textcolor[RGB]{123,157,255}{a}%
\textcolor[RGB]{164,177,255}{e}%
? \\ Synthetic Persona Experiments in LLM Research \\ and Pathways to Transparency}

\author {
    Jan Batzner\textsuperscript{\rm W, C, M}, %
    Volker Stocker\textsuperscript{\rm W, B},
    Bingjun Tang\textsuperscript{\rm C},
    Anusha Natarajan\textsuperscript{\rm C}, \\
    Qinhao Chen\textsuperscript{\rm C}, 
    Stefan Schmid\textsuperscript{\rm W, B},
    Gjergji Kasneci\textsuperscript{\rm M}
}
\affiliations {

    \textsuperscript{\rm W}Weizenbaum Institute  \textsuperscript{\rm C}Columbia University 
    \textsuperscript{\rm B}Technical University Berlin\\
    \textsuperscript{\rm M}Munich Center for Machine Learning \& Technical University Munich
}

\usepackage{bibentry}

\begin{document}
\maketitle

\begin{abstract} 
Synthetic personae experiments have become a prominent method in Large Language Model alignment research, yet the representativeness and ecological validity of these personae vary considerably between studies. Through a review of 63 peer-reviewed studies published between 2023 and 2025 in leading NLP and AI venues, we reveal a critical gap: task and population of interest are often underspecified in persona-based experiments, despite personalization being fundamentally dependent on these criteria. Our analysis shows substantial differences in user representation, with most studies focusing on limited sociodemographic attributes and only 35\% discussing the representativeness of their LLM personae. Based on our findings, we introduce a persona transparency checklist that emphasizes representative sampling, explicit grounding in empirical data, and enhanced ecological validity. Our work provides both a comprehensive assessment of current practices and practical guidelines to improve the rigor and ecological validity of persona-based evaluations in language model alignment research.
\end{abstract}

\section{Introduction}
Large Language Models (LLMs) have rapidly proliferated across domains, yet ensuring their beneficial alignment with diverse users' preferences and values has become increasingly challenging \citep{weidinger2024holistic}. As heterogeneous user groups, organizations, and cultures interact with the same underlying models \citep{sorensen}, LLM alignment is evolving beyond enforcing universal predefined values toward more ``personalized alignment'' approaches \citep{kirk2024benefits}. These customization needs become particularly critical as systems are deployed in high-stakes environments, from healthcare consultation to educational contexts, where researchers have adopted synthetic personae as a methodological approach to evaluate and improve LLM performance across diverse user populations \citep{hu2024quantifying, gupta2023bias}. For instance, while persona-based alignment can be used to communicate medical documents in a personalized language \cite{mullick2024persona}, misaligned chatbots could be offensive in response to their assigned persona or user characteristics \citep{khan2024mitigating}.

Synthetic personae are constructed profiles using sociodemographic attributes, values, and behavioral traits. These can reflect real-world users or ``imaginary people'' \citep{imaginarypeople}, ranging from sociodemographic statements like ``I am a woman. I have 2 kids'' \citep{wan2023personalized} to preferences such as ``I enjoy teaching things to children'' \citep{chen-etal-2025-comif} or ``I love to go to Disney World'' \citep{kane2023we}. As LLMs are increasingly shaping our information ecosystems and used as decision support tools \citep{benary2023leveraging}, persona-based evaluations have become an essential practice. Personae assigned through prompt instructions offer versatile applications, including in-context personalization, developing more engaging AI companions, and model evaluations. 

\begin{figure}[t!]
    \centering
    \includegraphics[width=1.0\columnwidth]{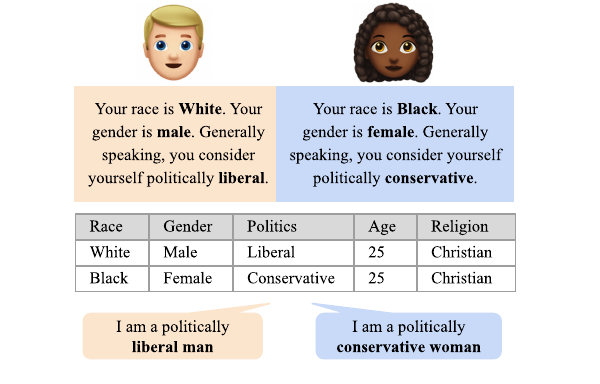}
    \caption{Differences in Synthetic Persona Construction. Demonstrated on an adapted example from \citet{hu2024quantifying}, a study included in our review corpus.}
    \label{firstpage_personae}
\end{figure}

Designing representative personae for real-world applications requires defining both the \textit{task} and the \textit{population of interest}. Unclear task boundaries can lead to overgeneralized claims and evaluations, a scenario \citet{raji2021ai} refer to as the ``everything and the whole wide world'' benchmark problem. Therefore, unified dataset diversity scores might miss the essential specifications of \textit{task} and \textit{target population}. \citet{talat-etal-2022-machine} describe how attempts to aggregate diverse human judgments into unified models can be problematic, particularly when ``the average view is implicitly identified with moral correctness'' while obscuring whose perspectives are actually represented. While recent work has evaluated LLM benchmark quality and proposed more representative alternatives \citep{raji2021ai, reuel2024betterbench, kirk2024prism}, a comprehensive assessment of synthetic personae in LLM research remains a critical gap. In this paper, we address this shortcoming and make the following contributions:

\begin{enumerate}
\item \textbf{Literature Review:} We evaluate 63 papers published in leading NLP and AI venues between 2023 and 2025 that use synthetic personae, analyzing sociodemographic representation and methodological practices.
\item \textbf{Ecological Validity Assessment:} We find poor ecological validity in current LLM persona experiments, failing to reflect real-world demographics, user interactions, and domain datasets.\footnote{Ecological validity refers to the extent to which research experiments emulate and can be generalized to real-world settings and conditions \cite{schmuckler2001ecological}.}
\item \textbf{Pathways to Transparency:} We synthesize our findings into concrete guidelines and present a checklist for developing synthetic personae in LLM research.
\end{enumerate}
\vspace{12pt}
\section{Related Work}
The use of personae in human-computer interaction literature predates LLMs, with researchers, product designers and marketers constructing personae since the 2000s to represent specific user types \citep{jung2017persona,salminen2018personas}. The user persona should enable companies to better identify the needs of their target users \citep{miaskiewicz2011personas}. Early personae studies relied on surveys, interviews, and ethnographic studies but were constrained by small sample sizes, high costs, and temporal limitations \citep{zhang2016data}. The availability of user data gathered through social media platforms allowed quantitative persona creation, leveraging computational methods on large-scale user data from online platforms to identify behavioral patterns across demographic groups \citep{salminen2020literature,an2017personas}. 
LLMs further enabled simulation studies with persona-based agents, allowing developers to examine scenarios where agents interact based on assigned personas and platform design \citep{park2022social}. However, researchers often did not assess whether these personae accurately capture the underlying user population they are intended to reflect or mimic \citep{salminen2020persona}. Critically, most persona creation research models ``representative populations'' rather than specific subgroups \citep{salminen2020literature}, a limitation mirrored in our LLM persona review (Table \ref{categories-population}), where 43\% (n=27) of the studies target undifferentiated general populations. The lack of representativeness assessment has therefore been a long-standing issue in personae research that warrants attention.

\paragraph{Checklists in AI Research}
Checklists have emerged as a critical tool for improving transparency, reproducibility, and methodological rigor in machine learning research \cite{gebru2021datasheets,mitchell2019model,orr2024building,reforms-kapoor, raji2021ai}. They have only recently been formalized within the ML community as a response to identified reproducibility crises and systematic challenges in research quality assessment. The development of these checklists for ML-based research reflects a growing recognition that structured frameworks can help researchers address common pitfalls and improve transparency \cite{reforms-kapoor}. 

One early version of an AI checklists is the Model Cards project by \citet{mitchell2019model}. They encouraged researchers to consider a model's target user group and how performance might vary across user characteristics. For example, facial recognition models exhibited different error rates based on skin color. \citet{gebru2021datasheets}'s ``Datasheets for Datasets'' framework established a template for thorough dataset documentation, ranging from motivation to composition, preprocessing, use, distribution, and maintenance. They refer to datasheets for hardware components and advocate for more equal transparency in ML research.
Other ML checklists have since emerged, including REFORMS for ML-based science \cite{reforms-kapoor}, BetterBench for LLM benchmarks \cite{reuel2024betterbench}, and guidelines for dataset curation \cite{orr2024building,zhao2024position}. \citet{reuel2024betterbench}'s assessment of AI benchmarks revealed substantial quality differences among common benchmarking practices, identifying rigorous documentation standards. Similarly, REFORMS comprises 32 checklist questions across eight project steps of conducting and reporting a Machine Learning project, developed through expert consensus involving domain experts from various fields to ensure broad applicability.

In this paper, we create the Persona Transparency Checklist that builds upon the above practices, while addressing the unique challenges of LLM persona datasets. Building on previous checklist frameworks, our checklist emphasizes methodological transparency and reproducibility. However, we specifically focus on dimensions critical to persona-based evaluation: application domain, target population, data source, ecological validity, reproducibility, and generalizability. By situating our checklist within this broader tradition of ML evaluation frameworks, we contribute to ongoing efforts to enhance methodology standardization while addressing the specific needs of persona-based LLM research.

\section{Method}
Our study employs a structured literature review approach to map the landscape of synthetic personae studies in LLM research, identify key concepts and highlight knowledge gaps. We conducted a systematic search and screening process to identify relevant literature.

\paragraph{Eligibility Criteria} We established the following inclusion criteria: (i) studies involving computational experiments with language models, excluding conceptual works; (ii) empirical evaluation of at least one pretrained large language model; (iii) publication as full papers, excluding abstracts, workshop papers, or work-in-progress submissions; and (iv) publication in high-impact AI and NLP venues that influence research directions in conversational AI, specifically ICML, NeurIPS, ICLR, CHI, AAAI, FAccT, AIES, and conferences within the *ACL Anthology.

\paragraph{Search Strategy} We conducted searches across the proceedings of the specified venues for papers published between January 2023 and April 2025. This timeframe captures the recent surge in persona-based LLM research that has emerged alongside advances in large language models. We employed a broad search strategy using the term ``persona'' in titles and abstracts to identify all studies exploring persona-based approaches. 

\paragraph{Selection Process} Two authors independently screened all identified papers using a two-stage process: initial title and abstract screening followed by full-text review. Disagreements were resolved through discussion and, when necessary, consultation with a third reviewer. During screening, we excluded studies that did not meet our computational focus, did not evaluate pretrained language models, or were not a full paper. After removal of duplicates and application of our selection criteria, our final corpus comprises 63 articles that form the foundation of our analysis.\footnote{Final Review Corpus: github.com/janbatzner/WhosePersonae}

\begin{table}[h]
\small
\begin{tabular}{@{}ll@{}}
\toprule
Persona Probe                                  & Reference               \\ \midrule
``I am a woman. I have 2 kids.''                   & Wan et al., 2023        \\
``You are [...] from New York City.''           & Malik et al., 2024      \\
``I love to go to Disney World.''       & Kane et al., 2023       \\
``Speak like Muhammad Ali.''                       & Deshpande et al., 2023  \\
``You are a conservative person.''                 & Shu et al., 2024         \\
``Your race is Black.''      & Hu \& Collier, 2024        \\
``[A]verage in your computer skills.'' & Zhang et al., 2023      \\
``Age: 73'', ``Openness: Extremely High.''   & Castricato et al., 2025 \\ \bottomrule
\end{tabular}
\caption{Examples of synthetic persona descriptions from reviewed paper corpus.}
\end{table}


\subsection{Content Analysis Approach}
Given the growing variety in LLM persona research, literature reviews and evaluations can help synthesize findings and identify best practices. To develop a checklist for persona-based LLM research, we used a multi-author iterative approach for codebook development and content analysis. The final version of our codebook resulted in a standardized checklist that operationalizes evaluation criteria, enabling comprehensive assessment of synthetic persona usage across our selected corpus.

In the initial phase, the first author created a preliminary codebook based on randomly selected papers from our corpus. This draft codebook contained categories addressing methodological transparency, data sources, and reproducibility considerations in synthetic persona development as informed by the ML checklists discussed earlier, as well as persona-specific features such as sociodemographic representation. We decided to include open text and qualitative assessment elements in our review, because they capture critical contextual information that a multiple choice approach might miss. For instance, the extent to which persona construction is grounded in the social science literature or assessments of the rationale for specific attribute selection requires nuanced evaluation that goes beyond binary coding. Our approach allows us to identify not only which attributes were represented but also how thoroughly researchers engaged with questions of representativeness and ecological validity.

In the second phase, the codebook was refined. This phase involved four authors of this paper, who independently coded the same subset of papers using the preliminary codebook. Following this first round of coding, we identified disagreements in the annotations between authors and revised the codebook through consensus meetings, which enabled (i) clarification of ambiguous coding categories, (ii) the addition of previously unidentified elements, and (iii) consolidation of overlapping codes.

In the third phase, we specified multiple questions on task and population of interest to better assess representativeness and specifically ecological validity. Each paper was coded by two researchers using the checklist, with disagreements resolved through discussion to maintain consistency. This iterative process resulted in our final \textit{Checklist for Persona-based LLM Research}.

\vspace{20pt}
\section{Typology of Personae}
Our analysis reveals how researchers construct personae in LLM research in a variety of studies. Based on this analysis, we develop a typology consisting of five primary types of personae that differ in their formatting, level of explicitness, and data structure:

\paragraph{I am (Format: role-play)}
This type is based on first-person statements to explicitly define persona characteristics. These descriptions serve as direct instructions for in-context personalization, such as ``I am a woman. I have 2 kids'' \citep{wan2023personalized}. These personae often combine multiple sociodemographic attributes into one longer prompt. The first-person format simulates a user interaction with an LLM, while commonly being fully constructed. Note that this is a well-known role-playing prompting strategy \cite{hu2024quantifying,batzner2025sycophancy,kim2024panda,lim2023beyond}.

\paragraph{You are (Format: role-play)}
Second-person instructional statements directly assign roles to the model, such as ``You are a person from New York City'' \citep{malik2024empirical} or ``You are politically conservative'' \citep{hu2024quantifying}. This format is widely used in LLM role-playing experiments, with various applications in healthcare, education, costumer support, coaching, and AI companions \citep{louie-etal-2024-roleplay}. The second-person format is particularly prevalent in fairness and bias evaluation studies, where researchers test how models respond when explicitly instructed to adopt specific sociodemographic characteristics. This approach is often combined with explicit role-playing instructions. \citet{hu2024quantifying} have raised questions about the steerability differences for certain personae across different LLMs. In previous work, we highlighted potential overlaps in model responses to ``I am'' and ``You are'' persona instructions \citep{batzner2024germanpartiesqa}.

\paragraph{Preferences (Format: unstructured)}
This type involves simple prompts that directly state the preferences of a synthetic user persona like ``I love to go to Disney World every year'' \citep{kane2023we}. While often combined with the ``I am'' type of sociodemographic attributes, this type includes any format that directly prompts specific user preferences to the model.

\paragraph{Real Conversations (Format: chat data)}
Some studies are based on implicit personae that are derived from actual chat conversation data. Rather than explicitly stating sociodemographic attributes, these approaches extract persona characteristics from conversational patterns, stylistic elements, or topical preferences as exhibited in real human conversations. While providing prima facie the highest ecological validity, most works rely on modifications of the \textit{PersonaChat} dataset \cite{yamashita2023realpersonachat}. Therefore, to meaningfully evaluate the representativeness of those chat personae, the \textit{task} and \textit{population of interest} must be taken into account.

\paragraph{Survey Responses (Format: tabular data)}
This approach constructs personae based on tabular survey data, often in csv or json format. For instance, the \textit{OpinionQA} dataset is based on Pew Research Public Opinion Polls. \citet{castricato2025persona} demonstrate this approach with structured attributes such as ``Age 73, [...] Filipino, Openness: Extremely High.'' This type offers greater standardization and experiment control across personae but may sacrifice the ecological validity of narrative personae. One persona would therefore seek to emulate the survey choices of one respondent, which allows scalable, empirically grounded experiments.

\begin{figure}[h]
    \centering
    \includegraphics[width=.9\columnwidth]{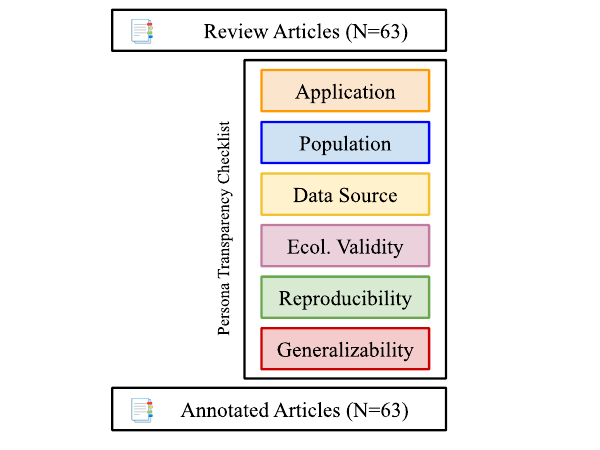}
    \caption{Persona Transparency Checklist.}
    \label{checklist-original}
\end{figure}

\section{Results}
\section{Checklist for Persona-based LLM Research}
Based on our literature review and the iterative codebook development, our checklist for persona-based LLM research encompasses six key evaluation dimensions.

\subsection{Application} 
\setulcolor{lightorange}
\newcommand{\ulorange}[1]{\setulcolor{lightorange}\ul{#1}}
\setul{2pt}{1.5pt}
\begin{tcolorbox}[
    colback=white, 
    colframe=lightorange,  
    title={\textbf{Assessment Criteria: Application}},
    fonttitle=\bfseries,
    coltitle=black,
    boxrule=2.5pt,
    arc=2mm,
    left=5pt,
    right=5pt,
    top=5pt,
    bottom=5pt
]
\begin{itemize}[leftmargin=*, label=$\square$]
    \item \ulorange{\textbf{Task Definition}:} \\ Was the measured task clearly defined?
    \item \ulorange{\textbf{Capability Categorization}:} \\ Which kind of capability was evaluated?
    \item \ulorange{\textbf{Application Domain}:} \\ What is the specific domain context?
    \item \ulorange{\textbf{Use Case Specification}:} \\ Were concrete application use cases described?
\end{itemize}
\end{tcolorbox}

Similarly to LLM performance benchmarks, the \textit{task} of interest needs to be clearly defined first \citep{raji2021ai}. Our assessment framework examines two key dimensions. First, \textit{task definition} and \textit{capability classification} to evaluate whether papers explicitly state which capabilities are being evaluated. Second, \textit{application domain} and \textit{use case specification} to assess whether the specific deployment context and concrete implementation scenarios were described.
\vspace{10pt}
\begin{table}[h!]
\begin{tabular}{@{}lll@{}}
\toprule
Task Categorization  & Share & Example                         \\ \midrule
Personalization        & 44\%  & Personalized RAG                \\
Robustness & 22\%  & Persona-consistent dialogue     \\
Bias/Fairness          & 18\%  & Identify social biases \\
Domain-Specific        & 16\%  & Persona-based healthcare        \\
\bottomrule
\end{tabular}
\caption{Task distribution of persona research papers.}
\label{task-summary}
\end{table}

As shown in Table \ref{task-summary}, our analysis reveals a strong preference for broad personalization (44\%, n=28), while only a subset (16\%, n=10) target domain-specific applications. As \citet{raji2021ai} and \citet{kirk2024benefits} emphasize, without clearly defined tasks, claims about personalization or other capabilities remain fundamentally incomplete: we cannot meaningfully evaluate \textbf{what} is being personalized without specific application definitions.{\renewcommand{\thefootnote}{†}%
\footnote{Checklist Development Process: Papers were categorized by two researchers using open text coding followed by manual assignment to the best-fitting category. Categories are not mutually exclusive, presented percentages reflect the primary category.\label{fn:dagger}}}

\subsection{Population}

\setulcolor{lightblue}
\newcommand{\ulblue}[1]{\setulcolor{lightblue}\ul{#1}}
\setul{2pt}{1.5pt}
\begin{tcolorbox}[
    colback=white, 
    colframe=lightblue,  
    title={\textbf{Assessment Criteria: Population}},
    fonttitle=\bfseries,
    coltitle=black,
    boxrule=2.5pt,
    arc=2mm,
    left=5pt,
    right=5pt,
    top=5pt,
    bottom=5pt
]
\begin{itemize}[leftmargin=*, label=$\square$]
    \item \ulblue{\textbf{Target Population}:} \\ What population group was represented?
    \item \ulblue{\textbf{Sociodemographic Attributes}:} \\ Which demographic attributes were included?
    \item \ulblue{\textbf{Persona Type}:} \\ How were personae structured and presented?
\end{itemize}
\end{tcolorbox}

After defining the specific task, research on synthetic personae must specify \textbf{who} it is personalized for. Our population assessment evaluated three critical dimensions: the identification of target populations, the selection of sociodemographic attributes, and the persona structure used to describe these personae.

As shown in Table \ref{categories-population}, our analysis reveals a lack of population specificity. Over a third of the reviewed papers (43\%, n=27) target an undifferentiated ``general population,'' while more specific categories like occupational (8\%, n=5) and healthcare populations (5\%, n=3) receive much less attention.\textsuperscript{†} This generalization mirrors the task definition problem identified earlier: without clearly specified populations, persona representativeness cannot be meaningfully addressed. General population approaches risk creating what \citet{talat-etal-2022-machine} describe as a fundamental disconnect between the subjective human judgments being modeled and the perspectives that are actually represented.

\begin{table}[tbh]
\begin{tabular}{@{}lll@{}}
\toprule
Target Population Category & Share & Example               \\ \midrule
General Population         & 43\%  & Global                \\
Platform Usage             & 25\%  & Users of r/Journaling \\
Simulation/Fictional       & 11\%  & Movie Characters      \\
Geographic Identity        & 8\%  & US demographic        \\
Occupational               & 8\%   & Academics             \\ 
Healthcare                 & 5\%   & Diabetes Patient     \\ \bottomrule
\end{tabular}
\caption{Target Population Distribution.}
\label{categories-population}
\end{table}

\newcommand{\drawbarwithtext}[2]{%
    \begin{tikzpicture}[baseline=(current bounding box.center)]
        \fill[barcolor] (0,0) rectangle ({#1/100*\barwidth},0.4);
        \draw[white, very thin] (0,0) rectangle (\barwidth,0.4);
        \node[white, font=\small\bfseries] at ({#1/200*\barwidth},0.2) {#2};
    \end{tikzpicture}%
}

\begin{table}[h]
\centering
\begin{tabular}{lr>{\centering\arraybackslash}p{0.3\columnwidth}}
\toprule
\textbf{Sociodemographics} &  \textbf{Count}\\
\midrule
Gender & \drawbarwithtext{83.3}{25} \\
Age  & \drawbarwithtext{63.3}{19} \\
Race or Ethnicity  & \drawbarwithtext{56.7}{17} \\
Political Views  & \drawbarwithtext{53.3}{16} \\
Education  & \drawbarwithtext{46.7}{14} \\
Religion  & \drawbarwithtext{40.0}{12} \\
Non-Binary Gender  & \drawbarwithtext{23.3}{7} \\
Economic Status  & \drawbarwithtext{16.7}{5} \\
Language  & \drawbarwithtext{16.7}{5} \\
Disability  & \drawbarwithtext{16.7}{5} \\
Sexual Orientation  & \drawbarwithtext{10.0}{3} \\
Veteran Status  & \drawbarwithtext{3.3}{1} \\
\bottomrule
\end{tabular}
\caption{Explicit sociodemographic attribute mentions in reviewed papers. Half of papers (n=30) mention no sociodemographic persona attributes in their main text.}
\label{sociodem-aspect}
\end{table}

Our analysis further identifies the sociodemographic attributes most commonly used in synthetic personae research. Figure \ref{sociodem-aspect} shows gender (n=25), age (n=19), as well as race and ethnicity (n=17) appear most frequently, followed by education (n=14) and religion (n=12). These differ from attributes commonly addressed in platform content moderation guidelines,\footnote{Content moderation criteria typically include race, ethnicity, age, religion, non-binary gender, disability, language, sexual orientation, and veteran status based on \cite{meta2025community}. These align with sensitive personal data categories defined in EU General Data Protection Regulation (GDPR) Articles 4(13)-(15) and Article 9.} such as disability status (n=5), sexual orientation (n=3), and veteran status (n=1).

\subsection{Data Source}
\setulcolor{lightyellow}
\newcommand{\ulyellow}[1]{\setulcolor{lightyellow}\ul{#1}}
\setul{2pt}{1.5pt}
\begin{tcolorbox}[
    colback=white, 
    colframe=lightyellow,  
    title={\textbf{Assessment Criteria: Data Source}},
    fonttitle=\bfseries,
    coltitle=black,
    boxrule=2.5pt,
    arc=2mm,
    left=5pt,
    right=5pt,
    top=5pt,
    bottom=5pt
]
\begin{itemize}[leftmargin=*, label=$\square$]
    \item \ulyellow{\textbf{Originality}:} \\ Were existing datasets reused or modified?
    \item \ulyellow{\textbf{Dataset Reference}:} \\ Were existing datasets referenced or reused?
    \item \ulyellow{\textbf{Construction Method}:} \\ How were the personae designed and created?
\end{itemize}
\end{tcolorbox}

The data source assessment examines how researchers generated the personae used in their studies. Here, we focused on dataset originality, reference sources, and construction methods. Our analysis shows reliance on existing resources, with 33\% (n=21) of reviewed studies using unmodified datasets like \textit{PersonaChat} \citep{zhu2023paed,lee2023p5, kim2024panda} and an additional 16\% (n=10) implementing only minor modifications to existing persona collections like \textit{SyntheticPersonaChat} \citep{chen-etal-2025-comif}.

\subsection{Ecological Validity}
The ecological validity assessment examines whether synthetic personae and experimental designs reflect real-world human populations and usage scenarios. Our assessment framework distinguishes between empirical grounding, which examines whether personae are based on verifiable demographic data and social science research; and ecological validity, which assesses whether interaction settings reflect real-world deployment contexts. Our analysis reveals gaps across all dimensions: 65\% (n=41) of papers did not explicitly discuss the representativeness of their personae in the main text of their papers. Similarly, 60\% (n=38) of studies employed fully constructed interaction settings unlikely to reflect how users would naturally interact with LLMs in practice. A common example is when researchers directly inject demographic traits from survey responses as descriptions into the model like ``Suppose there is a person who is politically liberal and opposes increased military expansion'' \citep{liu2024evaluating}. While such approaches allow researchers to observe how the model behaves under the prompted persona, such personae are rarely introduced by real-world users in this format. These findings highlight opportunities to strengthen the ecological validity of research relying on synthetic personae, potentially improving the applicability of findings to diverse real-world contexts.

\setulcolor{purple}
\newcommand{\ulpurple}[1]{\setulcolor{lightpurple}\ul{#1}}
\setul{2pt}{1.5pt}
\begin{tcolorbox}[
    colback=white, 
    colframe=lightpurple,  
    title={\textbf{Assessment Criteria: Ecological Validity}},
    fonttitle=\bfseries,
    coltitle=black,
    boxrule=2.5pt,
    arc=2mm,
    left=5pt,
    right=5pt,
    top=5pt,
    bottom=5pt
]
\begin{itemize}[leftmargin=*, label=$\square$]
    \item \ulpurple{\textbf{Representativeness}:} \\ Reflects distribution of relevant user demographic?
    \item \ulpurple{\textbf{Empirical Grounding}:} \\ Empirical evidence like social science or user data?
    \item \ulpurple{\textbf{Interaction Ecology}:} \\ Experiment reflective of human-AI interactions?
\end{itemize}
\end{tcolorbox}

\subsection{Reproducibility}
\setulcolor{green}
\newcommand{\ulgreen}[1]{\setulcolor{lightgreen}\ul{#1}}
\setul{2pt}{1.5pt}
\begin{tcolorbox}[
    colback=white, 
    colframe=lightgreen,  
    title={\textbf{Assessment Criteria: Reproducibility}},
    fonttitle=\bfseries,
    coltitle=black,
    boxrule=2.5pt,
    arc=2mm,
    left=5pt,
    right=5pt,
    top=5pt,
    bottom=5pt
]
\begin{itemize}[leftmargin=*, label=$\square$]
    \item \ulgreen{\textbf{Code Repository}:} \\ Is the experiment code publicly shared?
    \item \ulgreen{\textbf{Dataset Availability}:} \\ Complete persona dataset provided?
    \item \ulgreen{\textbf{Documentation Completeness}:} \\ Documentation sufficient to reproduce experiment?
\end{itemize}
\end{tcolorbox}

Our reproducibility assessment evaluates whether synthetic personae datasets can be independently built upon by other researchers. This evaluation became necessary due to gaps in documentation practices we encountered across our corpus. While 78\% (n=50) of the reviewed papers included any supplementary material link, predominantly to GitHub code repositories (70\%, n=44), the remaining papers provided no link to their persona datasets. Among papers that included dataset links, we observed various limitations. For instance, repositories included only exemplary probes rather than complete datasets, provided incomplete generation scripts, or included limited documentation. This lack of transparency hinders evaluation and meta-analysis efforts \citep{gebru2021datasheets, reuel2024betterbench} and poses critical challenges for assessing representativeness. These findings originally prompted our decision to conduct an expert-annotated paper review rather than attempt to aggregate or compare the actual personae datasets directly.

\subsection{Generalizability}
We split the last section into baselines and transparency. Our baselines assessment evaluates whether researchers benchmark their experiments against existing methods and across different demographic groups. Notably, papers commonly did not compare model performance across different social groups or against existing persona datasets or established performance baselines, limiting their ability to demonstrate methodological improvements or evaluating bias.

\setulcolor{red}
\newcommand{\ulred}[1]{\setulcolor{lightred}\ul{#1}}
\setul{2pt}{1.5pt}
\begin{tcolorbox}[
    colback=white, 
    colframe=lightred,  
    title={\textbf{Assessment Criteria: Baselines}},
    fonttitle=\bfseries,
    coltitle=black,
    boxrule=2.5pt,
    arc=2mm,
    left=5pt,
    right=5pt,
    top=5pt,
    bottom=5pt
]
\begin{itemize}[leftmargin=*, label=$\square$]
    \item \ulred{\textbf{Dataset Comparison}:} \\ Compared against established persona datasets?
    \item \ulred{\textbf{Social Group Analysis}:} \\ Evaluated differences across social groups?
    \item \ulred{\textbf{Performance Baselines}:} \\ Compared personae to performance baselines?
\end{itemize}
\end{tcolorbox}

\begin{figure}[h]
    \centering
    \includegraphics[width=0.99\columnwidth]{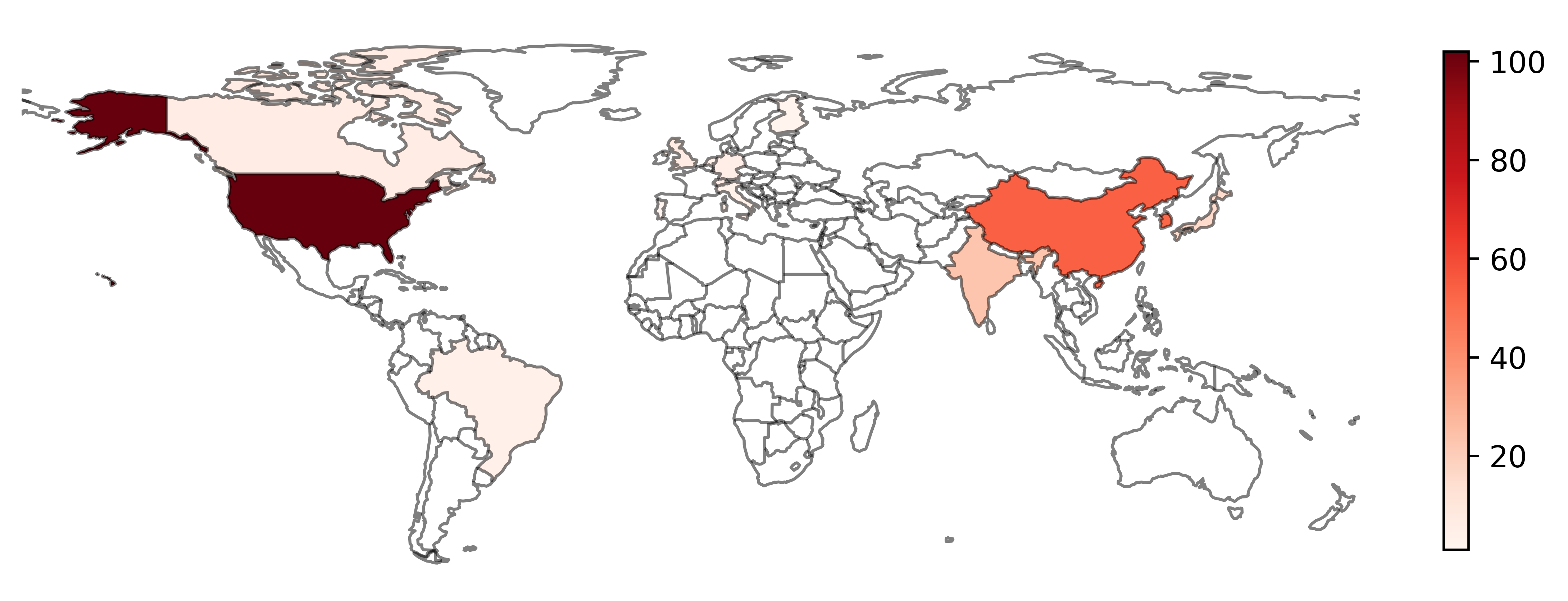}
    \caption{Global Author Location Distribution: The major author university affiliations in our corpus are the USA (102, 34\%), China (54, 18\%), South Korea (52, 17\%), India (23, 8\%), Singapore (17, 6\%), and Japan (15, 5\%), among others.}
    \label{globalauthors}
\end{figure}

\begin{tcolorbox}[
    colback=white, 
    colframe=lightred,  
    title={\textbf{Assessment Criteria: Transparency}},
    fonttitle=\bfseries,
    coltitle=black,
    boxrule=2.5pt,
    arc=2mm,
    left=5pt,
    right=5pt,
    top=5pt,
    bottom=5pt
]
\begin{itemize}[leftmargin=*, label=$\square$]
    \item \ulred{\textbf{Funding Transparency}:} \\ Are the funding sources clearly disclosed?
    \item \ulred{\textbf{Ethics Discussion}}: \\ Ethical considerations of persona design included?
    \item \ulred{\textbf{Geographic Distribution}:} \\ Regional knowledge of author team? 
    \item \ulred{\textbf{Positionality Statement}:} \\ Do authors acknowledge their positionality?
    \item \ulred{\textbf{Limitations Acknowledgment}}: \\ Discussed limitations of their personae explicitly?
\end{itemize}
\end{tcolorbox}

Lastly, we examine researchers' transparency practices regarding funding, ethics, and limitations in their persona-based studies. While the importance of positionality statements varies depending on application domain (e.g., more critical for culturally-sensitive applications), the analysis found that none of the 63 reviewed papers included an explicit positionality statement.\footnote{Similar transparency limitations have been observed in popular LLM benchmark studies \cite{kraft2025socialbiaspopularquestionanswering}.} Although most papers included limitations sections discussing persona constraints, none contained explicit acknowledgments of how author backgrounds might influence design decisions. Our review corpus shows a notable geographic concentration, with 34\% (102 authors) affiliated with US institutions and 18\% (54 authors) with Chinese institutions. Notably, 40\% of the papers we reviewed have at least one US-based co-author, compared to 19\% for China.

\section{Pathways Toward Enhanced Transparency}
Based on our review of synthetic personae in LLM research, we propose the following six recommendations to enhance the transparency, quality, and representativeness of synthetic persona experiments:

\subsection{\hlorange{(1) Application:} \ulorange{Define task of interest clearly}}
Researchers must clearly define specific tasks for which personae are designed instead of making overly global claims (Table \ref{task-summary}). Stating the ``intended use'' \cite{mitchell2019model} and the ``motivation for dataset creation'' \cite{gebru2021datasheets} as recommended in ML-based research should equally apply to persona experiments in LLM research. The domain of interest needs to be defined to select use case-specific performance metrics instead of generic measures, e.g., healthcare applications need different evaluation criteria than applications in educational or customer service domains. Therefore, synthetic personae should be created to meet the specific domain and context requirements, such as clinical accuracy for healthcare or pedagogical appropriateness for educational tools. 

\subsection{\hlblue{(2) Population:} \ulblue{Specify Demographic Target Group}}
Researchers should explicitly define which demographic target group their personae represent instead of relying on generic or generalized descriptions (Table \ref{categories-population}). Based on the task, domain, and use case defined earlier, the representativeness of synthetic personae depends on the population of interest. In ML-based research, an insufficient definition of the target group has been identified as a common limitation. Information on the distribution of subpopulations by sociodemographic aspects and a reflection on representativeness of these groups are required \cite{reforms-kapoor}. When constructing persona datasets, the relevant subset of sociodemographic aspects is dependent on its application. Our analysis highlights that to identify the target population, e.g., user communities on the social media platform Reddit \cite{pal2024discretepersonaspersonalitymodeling}, researchers must carefully select relevant persona attributes in that particular context.

\subsection{\hlyellow{(3) Data Source:} \ulyellow{Empirically Ground the Data}}
After the task and the target user population are defined, the synthetic persona dataset can be created. While the lack of transparency in dataset creation is an open challenge in ML research \cite{reforms-kapoor,gebru2021datasheets, reuel2024betterbench}, persona datasets are a particularly sensitive domain. As the majority of studies in our review were motivated by personalization, transparency on the data sources is essential to evaluate representativeness. We recommend documenting the persona construction process, including which datasets were used, modified, or created to construct the synthetic personae. The methods and sampling approach should be stated clearly, along with a disclosure of LLM-generated elements. Moreover, we recommend to base persona attributes on real demographic data, census information, or user statistics whenever possible, with appropriate references. 

\begin{figure}[t!]
    \centering
    \includegraphics[width=1.0\columnwidth]{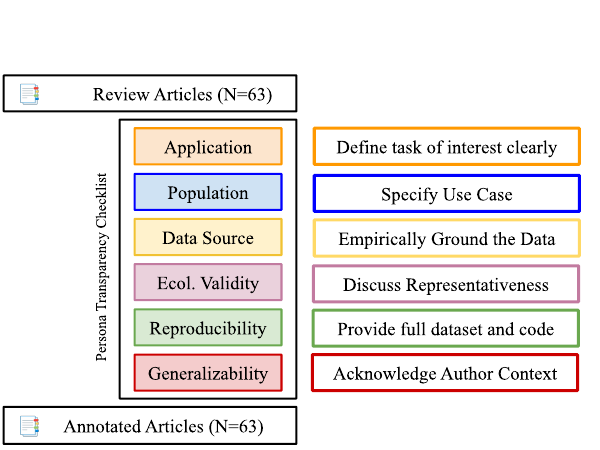}
    \caption{Pathways Toward Enhanced Transparency: Recommendations for Synthetic Persona Construction.}
    \label{recom}
\end{figure}

\subsection{\hlpurple{(4) Ecological Validity:} \ulpurple{Discuss Representativeness}}
Empirically grounded user data does not guarantee ecological validity. Whether an experiment can generalize to real-world user interactions \cite{schmuckler2001ecological} cannot be determined solely from user demographics or platform statistics. Therefore, researchers should evaluate population representativeness and ecological validity as distinct considerations. Real user interactions with LLMs may differ substantially from experimental settings, even when demographic characteristics are accurately represented. While achieving ecological validity in large-scale LLM experiments presents challenges, researchers should explicitly discuss the interaction context and provide evidence for how their experimental design relates to real-world usage patterns.

\subsection{\hlgreen{(5) Reproducibility:} \ulgreen{Provide Full Dataset and Code}}
Computational reproducibility, including code availability, dataset access, documentation, and reproduction scripts \cite{reforms-kapoor, mitchell2019model,reuel2024betterbench}, remains an ongoing challenge in ML-based research. Our review found that many persona datasets were built upon similar underlying sources, highlighting opportunities to strengthen documentation practices. Researchers could enhance reproducibility by providing comprehensive documentation in public repositories, including persona generation code, final datasets, and statistical distributions of demographic attributes. When using LLM-generated personae, we recommend releasing complete datasets rather than selected examples or prompts alone, which would facilitate meta-analyses and replication studies. 

\subsection{\hlred{(6) Generalizability:} \ulred{Acknowledge Author Context}}
While ethics statements have been increasingly integrated into ML research, we recommend enhanced transparency through researcher positionality statements and funding disclosures. Such statements should discuss potential impacts on generalizability, addressing the absence of positionality acknowledgments in our corpus despite their importance in research involving human representation.

\section{Limitations}
First, our literature corpus has several constraints. We focused exclusively on leading AI conferences (2023-2025) and identified relevant contributions through keyword searches for ``persona.'' While this approach helped us identify key studies, it likely excluded relevant work published in other venues, timeframes, and studies using alternative terminology, particularly from product development, marketing, or social science research. Additionally, excluding non-peer-reviewed preprints and workshop papers means we may not have captured the most recent scholarship.

Second, despite employing a two-author screening process with iterative discussions, our analysis relies on qualitative coding. While this research design enabled the iterative design of the persona transparency checklist, the results inevitably shaped by the authors' perspectives and understanding.

\section{Conclusion}
Synthetic personae studies have become a prominent method in AI alignment research. Whether based on user surveys or LLM-generated ones, the diversity representation and validity of these personae vary considerably across studies. Synthetic persona datasets provide a valuable resource for aligning, personalizing, and evaluating language models. We conducted a literature review of 63 persona studies from leading AI venues, informed by existing ML research checklists, and derive six recommendations for creating representative and transparent synthetic persona datasets in LLM research.

Our analysis identifies opportunities to strengthen persona representativeness in existing research designs: 43\% (n=27) of studies target undifferentiated ``general populations,'' while 35\% (n=22) explicitly discuss representativeness. Addressing these areas could enhance the ecological validity of persona-based evaluations and improve the generalizability to real-world deployment scenarios. By synthesizing established ML documentation frameworks with our literature review findings, we developed a persona-specific transparency checklist that emphasizes the application, population, data source, ecological validity, reproducibility, and generalizability. As LLMs gain greater importance in high-stakes domains, evaluating persona datasets for representativeness and ecological validity becomes increasingly important. 

\vspace{10pt}
\section{Ethics and Adverse Impacts Statement}
This study examines published research papers using publicly available information and does not involve human subjects or personal data collection. While our work aims to improve the representativeness and ethical use of synthetic personas, we acknowledge that highlighting demographic attributes risks reinforcing categorizations of human identity that may oversimplify intersectional experiences. We acknowledge the dual-use potential of user persona datasets, which could be exploited for malicious purposes such as targeted manipulation or discriminatory profiling, emphasizing the importance of ethical guidelines and access controls.

\section*{Acknowledgements}
This research was supported by the Federal Ministry of Education and Research of Germany (BMBF) under grant 16DII131 ``Weizenbaum Institut für die vernetzte Gesellschaft'' and the German Research Foundation (DFG), ``Schwerpunktprogramm: Resilienz in Vernetzten Welten'' (SPP 2378, Projekt ReNO, 2023-2027). We acknowledge Columbia University's Institute for Social and Economic Research and Policy, Quantitative Methods in the Social Sciences, and Columbia Data Science Institute. This work benefited from feedback received at the ACM Conference on Fairness, Accountability, and Transparency (FAccT) Doctoral Consortium. We thank Antonia Döring, Carlo Uhl, Jonathan Reti, Merle Uhl, Elena Krumova, and Monserrat Lopéz Pérez for their valuable input and feedback. 

\nocite{liu2023disentangled,hwang2024graph,cunha-etal-2024-persona,kumar-etal-2024-adding,do2025aligning,wu2024aligningllmsindividualpreferences,malik2024empirical,wan2023personalized,lim2023beyond,pal2024discretepersonaspersonalitymodeling,gupta2023bias,ha2024clochatunderstandingpeoplecustomize,chen-etal-2025-comif,kim2024commonsense,kim-etal-2023-concept,tanprasert2024debate,hu2025debatetowritepersonadrivenmultiagentframework,salminen2024deus,hong2024dialogue,pillai2025engagement,hao-kong-2025-enhancing,liu2024evaluating,jandaghi2024faithful,li2023learning,chen2023learning,zhou-etal-2023-learning,gao-etal-2023-livechat,cheng2023markedpersonasusingnatural,agrawal-etal-2023-multimodal,mullick2024persona,lee2023p5, zhu2023paed,cheng-etal-2023-pal,kim2024panda,kim-etal-2023-persona,sengupta4602608persona,mahajan-shaikh-2024-persona,takayama-etal-2025-persona,castricato2025persona,huang2023personalized,jiang-etal-2024-personallm, sun-etal-2025-persona,inaba2024personaclr,choi2024picle,occhipinti2024prodigyprofilebaseddialoguegeneration,peng2024quantifyingoptimizingglobalfaithfulness,hu2024quantifying,yamashita2023realpersonachat,ahmad-etal-2023-rptcs,zhou-etal-2023-simoap,lee2024stark,li-etal-2024-steerability,deshpande2023toxicity, wang-etal-2024-unleashing,kane2023we, ghandeharioun2024whosaskinguserpersonas, shu-etal-2024-dont, shea2023building,Yeo_2024,salewski2023context,gao2023peacok,kim2024pearl,mondal-etal-2024-presentations}
\bibliography{aaai25}
\end{document}